\documentclass[aps,prl,reprint,twocolumn,showpacs,floatfix,superscriptaddress]{revtex4-2}
\usepackage{amssymb,amsmath,amstext}                %%   American Physical Society math etc extensions
\usepackage{graphicx}%
\usepackage{dcolumn}%
\usepackage{bm}%
\usepackage{physics}
\usepackage{color}
\usepackage{ulem} %

\usepackage[colorlinks, linkcolor=blue, anchorcolor=blue, citecolor=blue]{hyperref}

\begin{document}
\title{Dynamical quantum phase transitions from quantum optics perspective}%
\author{Jakub Zakrzewski} 
\affiliation{Instytut Fizyki Teoretycznej, Wydzia\l{} Fizyki, Astronomii i Informatyki Stosowanej, 
Uniwersytet Jagiello\'nski,  ulica profesora Stanis\l{}awa \L{}ojasiewicza 11, PL-30-348 Krak\'ow, Poland}
\affiliation{Mark Kac Complex Systems Research Center, Uniwersytet Jagiello{\'n}ski, Krak{\'o}w, Poland}

\date{\today}

\begin{abstract}

In this work dedicated to Professor Iwo Bia\l{}ynicki-Birula on the occasion of his 90th birthday, I attempt to show that dynamical quantum phase transitions observed as singularities in the Loschmidt rate dynamics bear a close resemblance to standard Rabi oscillations known from the dynamics of two-level systems. For some many-body systems, this analogy may go even further, and the behaviour observed for example for transverse Ising chain can be directly mapped to such simple dynamics.   A simple link between Loschmidt echo singularities and quantum scars is also suggested.

\end{abstract}
\maketitle

\section{Introduction}
The physics of complex systems may sometimes be understood (in particular limiting cases) in a  simple, enlightening form. This has been  often demonstrated in quantum optics, one of the many areas of Iwo Bia\l{}ynicki-Birula outstanding contributions. As a scientific grandson of Iwo Bia\l{}ynicki-Birula, I had relatively small overlap in scientific interests with him, our paths crossed, for a moment,  in the studies of  nonspreading wave-packets \cite{Bialynicki94,Delande_1995,Delande98,Buchleitner02}. Still, however,
I profited a lot from occasional conversations as well as participation, from time to time, in unusually vivid seminars with his active participation. Often his aim was to find a simple picture of the presented effects.
In this contribution I consider briefly two cases from studies 
of nonequilibrium dynamics of many body systems which may be, in my opinion, understood in simple terms:  the dynamical quantum phase transitions (DQPT) \cite{Heyl13,Heyl18} and quantum many-body scars (QMBS)  dynamics \cite{Bernien17}.

The simplest definition of DQPT consists of a sudden quench in which the system is prepared in thr ground
state $\Psi\rangle$ of a parameter dependent Hamiltonian $H(\lambda=0)$, and $\lambda$ is suddenly changed to other
value. It has been observed that often, if the change of $\lambda$ moves the Hamiltonian into a different phase, the time dynamics with $H(\lambda)$ of  the now nonstationary state after the quench reveals the so called Loschmidt echo singularities.
Their appearance is neither a necessary nor a sufficient condition for the phase transition between $H(0)$ and $H(\lambda)$. Still predominantly lack of singularities occurs if no phase transition is crossed while changing $\lambda$ and vice versa.

Dynamical detection of quantum scars \cite{Bernien17} is in some sense similar. One prepares an initial nonstationary state for a many-body system described by $H(\lambda)$ . When this initial state has a significant overlap with  a few almost equally spaced in energy eigenstates of   $H(\lambda)$ - the time evolution of observables reveals oscillations even in a weakly ergodic regime, i.e. when the dynamics of a typical generic state will lead to thermalization.

Both these phenomena, while of current interest, can be simply explained by indentifying the ``essential state model'' i.e. a minimal approximate  level  scheme allowing one to simulate the dynamics. Let us consider first consider DQPT in the seminal example of a transverse Ising model.

\section{DQPT in transverse Ising model} 

The first work on DQPT \cite{Heyl13} considers the transverse Ising model with Hamiltonian of the form
\begin{equation}
H=-\frac{1}{2}\sum_i \sigma^z_i\sigma^z_{i+1} -\frac{g}{2}\sum_i \sigma^x_i
\label{TIM}
\end{equation}
where $g$ is the strength of the magnetic field pointing in $Ox$ direction. For small $g$ the interactions favor ferromagnetic (FM) (along $Oz$) orientation, with two degenerate
(in the thermodynamic limit) ground states given for $g\rightarrow 0$ by $\ket{\psi(\pm)^z} = \prod_i \ket{\pm}^z_i$, where $\ket{\pm}^z_i$ denotes eigenvectors of $\sigma^z_i$. The phase transition from FM to paramagnetic order occurs for $g=1$,
for large $g$, the unique ground state is well approximated by  $\ket{\varphi} =  \prod_i \ket{+}^x_i$ with $\ket{\pm}^x_i$ bering eigenvectors of $\sigma^x_i$.

Let $g$ serve as a parameter $\lambda$ and let us start with the ground state of \eqref{TIM} for small $g$, say with   $\ket{\Psi} = \ket{\psi(+)^z}$ and abruptly change $g$ to large, positive value. In the new Hamiltonian the term proportional to $g$ will dominate while the first interaction term will be a small perturbation. The initial state can be then decomposed in the basis of $\sigma^x_i$ eigenvectors as: 
\begin{equation}
 \ket{\Psi(0)}= \prod_i \frac{1}{\sqrt{2}} (\ket{+}^x_i + \ket{-}^x_i).
 \end{equation} 
  The initial state after a quench is, therefore, a product of two-state combinations with coefficients of equal magnitude (the phase does not affect the result). The subsequent time evolution, still neglecting interactions in the final Hamiltonian yields
  \begin{equation}
 \ket{\Psi(t)}= \prod_i \frac{1}{\sqrt{2}} \left(\ket{+}^x_i\exp(igt/2)  + \ket{-}^x_i\exp(-igt/2)\right).
 \end{equation} 
By a survival probability (fidelity, return amplitude, or Loschmidt echo) one calls (depending on the context) the squared overlap of initial and time evolved state, ${\cal L}(t) \equiv |\langle \Psi(0)\ket{\Psi(t)}|^2$. Further one may define \cite{Heyl18} the rate function, $r(t)$ via  ${\cal L}(t)=\exp{-L r(t)}$ where $L$ is the system size (number of degrees of freedom). Such a measure has a good thermodynamic limits. Singularities in $r(t)$ time dependence, often referred to as Loschmidt echo singularities, are the defining features of DQPT.

Let us immediately consider the simplified example above. The squared overlap ${\cal L}(t)$ becomes simply ${\cal L}(t)=\cos^{2L}(gt/2)$ and size-independent rate $r(t)$ reveals singularities whenever the cosine function vanishes i.e. for $t^*=(2k+1)\pi/g$ for integer $k$.  This example clearly shows that Rabi-type oscillations are the real origin of rate function 
singularities in this case. 

One can complain that the situation described above is too simplified, singularities in form of finite cusps appear also for smaller changes of $g$ where the approximations made by us
would not work fully. Then, however, one can use Jordan-Wigner transformation into the noninteracting fermion system, as in the original DQPT letter \cite{Heyl13} and observe similar ``two-level'' dynamics for a given $k$ as different $k$ decouple. 

\section{Other examples} 

Our model, however, helps to explain also other situations. In fact, as reviewed in \cite{Heyl18} 2-band topological noninteracting models lead to exactly to the same dynamics. Again here, due to lack of interactions, different $k$ values can be treated independently leading to similar estimate of critical times at which singularities appear. Let us stress that while these singularities are essential for phase-transition language application, they seems just to be due to vanishing overlaps between the initial and time evolved wavepacket.

Consider now the situation in which 
we make an abrupt quench within the same phase then by definition the ground state changes slowly and continuously with the change of the parameter for a finite system.
 So it is quite justified to assume that the ground state at say $\lambda=0$ expands, in eigenstates $\{\ket{\psi_k} \}$ of $H(\lambda)$ as
\begin{equation}
\ket{\Psi} = \alpha_0 \ket{\psi_0} + \sum_k \alpha_k \ket{\psi_k},
\end{equation}
with $|\alpha_0| \gg |\alpha_k$ for $k>0$.  Then the survival probability (Loschmidt echo) is dominated by a large $|\alpha_0|^2$ term. The situation is more subtle in the thermodynamic limit due to Anderson 
catastrophe. Still then one may expect that many eigenstates at final parameter value contribute to the initial wavepacket, leading to many superimposed oscillations at different
frequencies. In such a situation rate function should not reveal strong maxima (not speaking of singularities).

Note that situation is markedly different when the phase transition is crossed in $\lambda$ as then, for an Ising system, via symmetry, as described above, {\it two} eigenstates contribute significantly to the sum, leading to Rabi oscillations at half of their energy difference (per site).

The discussion up till now was concentrated on spin-1/2 models leading to simple Rabi oscillations. This might be a transverse Ising chain but also, e.g. a quantum dot dynamics \cite{Wrzesniewski22}.  As known from quantum optics, Rabi oscillations generalize  to quantum revivals appearing when several equally spaced levels are populated \cite{Eberly80}. Here again one may expect that {\it between} consecutive revivals minima of the survival probability lead to maxima (and possibly cusps) of the Loschmidt rate functions.  In a many body system, even a more general sitation was experimentally realized many years ago for interacting bosons in an optical lattice \cite{Greiner02r}. Initially bosons were kept in a shallow lattice, then, abruptly, the lattice height was dramatically increased separating different lattice sites. Within each site the initial, almost coherenty state was a superposition of states with different site occupations separated by the quadratic progression in the interaction strength $U$ (within the tight binding Bose-Hubbard description) and the revivals were observed. The corresponding Loschmidt rates reveal singularities (or maxima) as discussed in detail recently \cite{Lacki19} in the DQPT language at times when the overlap between initial and time evolved state is minimal i.e. roughly in the middle between two consecutive revivals.

\section{Quantum scars}

Recently an interesting manifestation of ergodicity breaking as persistent oscillations for certain initial states was discovered experimentally with ultracold Rydberg atoms \cite{Bernien17}. This feature is due to the presence of few atypical, almost equally spaced eigenstates  -- the so-called quantum many-body scars (\textbf{QMBS}) \cite{Turner18, Turner18q} that are embedded in the otherwise thermal spectrum of a quantum many-body system. For initial states with high overlap with a few QMBS, one observes long-lived oscillations of observables, whereas for generic initial conditions the system quickly approaches the thermal equilibrium state. The same oscillations should be present in the survival probability leading, in turn, to maxima of Loschmidt echo rate function if the data are interpretted in that way.

QMBS borrowed their name from single particle quantum chaos studies where ``quantum scar'' described enhanced probability of eigenstates or wavepackets in the regions of space occupied by unstable periodic orbits \cite{Heller84} - in close relation to semiclassical periodic orbits quantization \cite{Gutzwiller71,Bogomolny88}. Then also the concept of scarring by symetries were developed in the contex of hydrogen atom in magnetic field studies \cite{Delande87}. Similar symmetry concepts were used for nonergodic states construction in many-body case see, e.g. \cite{Shiraishi17,Moudgalya18}.

Such QMBS may be easily imagined as having the origin in approximate decoupling of a (not always apparent) single degree of freedom from the other degrees of freedom. If this single degree is locally described by 
an harmonic oscilator (or an angular momentum) then the corresponding eigenstates are equidistant - their weak coupling to  the remaining states preserves the energy structure. Now, if by accident (or cleverness) the initial state is prepared as a linear combination of those selected states (or if it has sufficiently large overlap on at least a few of them) one may naturally expect a persistent oscillation in the time dynamics. 
Let us mention also that quench dynamics and Rabi oscillations resulting from excitations of two or more localized integrals of motion in the context of many-body localization have been studied recently \cite{Benini21a,Benini21b}. The localized, almost decoupled family of states may not be easy to identify, one may try to identify it e.g., by adiabatic following from some analytic limit \cite{Adith22} or via purely numerical approaches including artificial intelligence \cite{Szoldra22}.

\section{Conclusions} 
DQPT form a very intriguing interpretation of rapid quantum quenches. On the other hand, the signatures of DQPT in the form of singularities of Loschmidt echo  rate functions, appear to a large extend due to the very definition of this rate. The survival probability (Loschmidt echo) itself reveals no singularities but rather smooth oscillations (or revivals in more complicated cases).

Let us stress that the mechanism presented above considers rather simple examples. For more complicated cases one may consider Loschmidt echo as coming back not to a single ground state but to the degenerate manifold if it exists \cite{Heyl18}. After the first draft of this note was completed,   a related work appeared giving  a more general picture of DQPT \cite{Damme22}. It has been brought also to our attention that similar to DQPT cusp structures may appear in single particle dynamics \cite{Zhang16,Zhang16b}.

\section{Acknowledgments}    
This note was born as a reflection on a seminar given in Krakow by Tadeusz Doma\'nski. His great presentation skills are acknowledged.
I am grateful to Dominique Delande, Mateusz \L{}\c{a}cki, Anatoly Polkovnikov, and Tommaso Roscilde for discussions.
The work of J.Z have been realized within the Opus grant
 2019/35/B/ST2/00034, financed by National Science Centre (Poland). The research  has also been supported by a grant from the Priority Research Area DigiWorld under the Strategic Programme Excellence Initiative at Jagiellonian University.
 

%\bibliography{ref_21}

\end{document}